\documentclass[twocolumn]{aastex631}

\begin{document}

\title{A centi-pc-scale compact radio core in the nearby galaxy M60}

\correspondingauthor{Xiaofeng Li and Xiaopeng Cheng}
\email{lixf@gzhu.edu.cn, xcheng@kasi.re.kr}
\author[0000-0002-9093-6296]{Xiaofeng Li}
\affiliation{Department of Astronomy, Guangzhou University, Guangzhou 510006, PR China}
\affiliation{Great Bay Center, National Astronomical Data Center, Guangzhou, Guangdong 510006, China}
\affiliation{Astronomy Science and Technology Research Laboratory of Department of Education of Guangdong Province, Guangzhou 510006, China}

\author[0000-0002-2322-5232]{Jun Yang}
\affiliation{Department of Space, Earth and Environment, \\
Chalmers University of Technology, Onsala Space Observatory, SE-439 92 Onsala, Sweden}

\author[0000-0003-4407-9868]{Xiaopeng Cheng}
\affiliation{Korea Astronomy and Space Science Institute, Daedeok-daero 776, Yuseong-gu, Daejeon 34055, Republic of Korea}

\author[0000-0002-9137-7019]{Mai Liao}
\affiliation{National Astronomical Observatories, Chinese Academy of Sciences, 20A Datun Road, Chaoyang District, Beijing 100101, China}
\affiliation{Chinese Academy of Sciences South America Center for Astronomy, National Astronomical Observatories, CAS, Beijing, 100101, China}
\affiliation{Instituto de Estudios Astrofísicos Facultad de Ingeniería y Ciencias Universidad Diego Portales Av. Ejército 441, Santiago, Chile}

\author[0000-0002-1992-5260]{Xiaoyu Hong}
\affiliation{Shanghai Astronomical Observatory, Chinese Academy of Sciences, 200030 Shanghai, People’s Republic of China}
\affiliation{Shanghai Tech University, 100 Haike Road, Pudong, Shanghai, 201210, People’s Republic of China}
\affiliation{University of Chinese Academy of Sciences, 19A Yuquan Road, Beijing 100049, People’s Republic of China}

\author[0000-0002-4757-8622]{Liming Dou}
\affiliation{Department of Astronomy, Guangzhou University, Guangzhou 510006, PR China}
\affiliation{Great Bay Center, National Astronomical Data Center, Guangzhou, Guangdong 510006, China}
\affiliation{Astronomy Science and Technology Research Laboratory of Department of Education of Guangdong Province, Guangzhou 510006, China}

\author{Tianle Zhao}
\affiliation{Department of Astronomy, Guangzhou University, Guangzhou 510006, PR China}
\affiliation{Great Bay Center, National Astronomical Data Center, Guangzhou, Guangdong 510006, China}
\affiliation{Astronomy Science and Technology Research Laboratory of Department of Education of Guangdong Province, Guangzhou 510006, China}

\author{Zhongying Fan}
\affiliation{Department of Astronomy, Guangzhou University, Guangzhou 510006, PR China}
\affiliation{Great Bay Center, National Astronomical Data Center, Guangzhou, Guangdong 510006, China}
\affiliation{Astronomy Science and Technology Research Laboratory of Department of Education of Guangdong Province, Guangzhou 510006, China}

\author[0000-0002-0403-9522]{Fupeng Zhang}
\affiliation{Department of Astronomy, Guangzhou University, Guangzhou 510006, PR China}
\affiliation{Great Bay Center, National Astronomical Data Center, Guangzhou, Guangdong 510006, China}
\affiliation{Astronomy Science and Technology Research Laboratory of Department of Education of Guangdong Province, Guangzhou 510006, China}

\author[0000-0001-8449-6020]{Weirong Huang}
\affiliation{Department of Astronomy, Guangzhou University, Guangzhou 510006, PR China}
\affiliation{Great Bay Center, National Astronomical Data Center, Guangzhou, Guangdong 510006, China}
\affiliation{Astronomy Science and Technology Research Laboratory of Department of Education of Guangdong Province, Guangzhou 510006, China}


\submitjournal{ApJ}
\begin{abstract}
M60, an elliptical galaxy located 16.5~Mpc away, has an active nucleus with a very low luminosity and an extremely low accretion rate. Its central supermassive black hole has a mass of $M_{\rm BH}\sim4.5\times10^{9}\, M_{\odot}$ and a Schwarzschild radii corresponding to $R_{\rm S}\sim5.4\,\mu\mathrm{as}$. To investigate the nature of its innermost radio nucleus, data from the Very Long Baseline Array (VLBA) at 4.4 and 7.6~GHz were reduced. The VLBA images reveal a compact component with total flux densities of $\sim$20~mJy at both frequencies, a size of $\leq$0.27~mas (99.7$\%$ confidence level), about 0.022~pc ($50\,R_{\rm S}$) at 7.6~GHz, and a brightness temperature of $\geq6\times10^{9}$~K. This suggests that the observed centi-parsec-scale compact core could be attributed to a nonthermal jet base or an advection-dominated accretion flow (ADAF) with nonthermal electrons. The extremely compact structure also supports the presence of an SMBH in the center. Our results indicate that M60 is a promising target for broad-band VLBI observations at millimeter wavelengths to probe ADAF scenarios and tightly constrain the potential photon ring (about 28\,$\mu$as) around its SMBH.

\end{abstract}



\section{Introduction} \label{sec:intro}
Spectroscopy and dynamical modeling suggest that supermassive black holes (SMBHs) are usually located in the center of massive galaxies \citep{Gebhardt2000, Kormendy2013}. The most powerful evidence for the presence of an SMBH in the center of a galaxy is the detection of the inner radio structure close to the SMBH candidates, e.g., M87 \citep{EHTC2019}, Sagittarius A* \citep[Sgr A*,][]{EHTC2022}, NGC 4258 \citep{Miyoshi1995}. A spectroscopic survey of nearby bright galaxies found that around $40\%$ of them contain low-luminosity active galactic nuclei (LLAGNs) that are powered by SMBHs \citep{Ho1997, Ho2008}. Generally, the radiation of most LLAGNs originates from sub-pc-scale or even smaller regions, making it difficult for radio observations to resolve these regions. The radiation mechanism of LLAGNs is not yet well understood.

Radio images of LLAGNs often display a compact core, and some of them also have jet-like structures \citep{Ho2001, Nagar2005}. These radio cores remain compact on milliarcsecond scales, as observed through very long baseline interferometry \citep[VLBI, e.g., ][]{Ulvestad1999ApJL, Ulvestad2001, Anderson2004, Falcke2000, Nagar2002, Nagar2005, Doi2005, Park2017}. A few LLAGN have been resolved on scales of $\leq100 $ Schwarzschild radii, $R_{\rm S}$, including Sgr A* \citep{Shen2005, Lu2018, EHTC2022}, M87 \citep{Hada2011,EHTC2019,Lu2023}, Cen A \citep{Janssen2021}, M104 \citep{Hada2013}, M81 \citep{Jiang2018}, and M84 \citep{Jiang2021}. The radio cores of LLAGNs typically have a flat or slightly inverted spectrum \citep{Ho2001, Nagar2005, Anderson2004, Cho2022}. The LLAGN jets usually have subluminal speeds \citep{Ulvestad1999ApJL, Hada2013}.

The radio emission of LLAGNs has multiple origins, one of the most common being outflows. These outflows consist of collimated jets \citep{Blandford1977, Falcke1999, Chen2021}, accelerated to relativistic velocities \citep{Lister2021} with a very small opening angle \citep{Pushkarev2017}, and winds \citep{Blandford1982, Blandford1999} with low brightness, a large opening angle and low velocity \citep{Panessa2019}. The jets emit a power law continuum spectrum in the radio band, which is caused by nonthermal synchrotron radiation. Radio winds have the same radiation mechanism as jets, but are much weaker than jets due to low energy and weak magnetic fields. They have only been detected in a few sources, e.g., Markarian 6 \citep{Kharb2006, Kharb2014}. Additionally, advection-dominated accretion flows (ADAFs), characterized by geometry thick and optical thin, also emit significant radio flux \citep{Narayan1994, Narayan1995A, Yuan2003}. The radio emission from ADAF primarily dominates in the very inner region, making it difficult to study with low-resolution observations \citep{Mahadevan1997, Manmoto1997, Bandyopadhyay2019}. Lastly, star formation activity and supernova remnants can also contribute significant radio emission, adding complexity to the study of outflow and ADAF \citep{Panessa2019}. 

Sgr A*, the closest known SMBH at a distance D $\sim$8.1 kpc \citep{GRAVITYCollaboration2019}, is one of the most promising targets for studying the vicinity of SMBHs in extremely low accretion rate \citep{EHTC2022}.
Despite extensive VLBI observations, no direct evidence of a jet has been found in Sgr A* \citep{Rauch2016, EHTC2022, Cho2022, Cheng2023}. Interstellar scattering \citep{Bower2006, Psaltis2018} is a major obstacle to detecting potential jets in Sgr A*. To avoid complications caused by interstellar scattering, we conducted a comprehensive literature review \citep[e.g., ][]{Kormendy2013, Ramakrishnan2023} and identified M60 as a prime candidate for studying the presence of jets in LLAGNs with very low luminosity and extremely low accretion rates.

M60 (NGC 4649), a nearby elliptical galaxy, has an active nucleus \citep{DiMatteo1997, DiMatteo1999, Paggi2017}. Its high Galactic latitude ($b=74.3^\circ$) implies that the interstellar medium (ISM) has a negligible effect on scatter broadening \citep{Koryukova2022}. The distance to M60, as estimated by the tip of the red giant branch (TRGB) method, is $d=16.23\pm0.50({\rm ran})\pm0.42({\rm sys})~$Mpc \citep{Lee2017}, and the surface brightness fluctuation (SBF) method gives an estimate of $d=16.7\pm0.6~$Mpc \citep{Cantiello2018}. In this work, we use the weighted average of the two distances, $d = 16.5\pm0.5~$Mpc. The image scale of M60 is $0.080\pm0.003~{\rm pc\,mas^{-1}}$. The central object, most likely an SMBH, has a dynamic mass of  $M_{\rm BH} = (4.5\pm1.0)\times10^{9}~M_{\odot}$ \citep[scaled to solar mass, black hole mass represented by $m=(4.5\pm1.0)\times10^{9}$,][]{Shen2010}. The central SMBH candidate has an angular size of $R_{\rm S}=2GM_{\rm BH}/(c^{2}d)\sim 5.4~{\rm \mu{}as}$, where $G$ and $c$ are the gravitational constant and the speed of light, respectively. The mass accretion rate of an SMBH $\dot{M}$ is often scaled to Eddington rate $\dot{M}_{\rm Edd}=10L_{\rm Edd}/c^{2}$, i.e.,  $\dot{m}=\frac{\dot{M}}{\dot{M}_{\rm Edd}}$, where $L_{\rm Edd}$ is the Eddington luminosity\footnote{The Eddington luminosity is the maximum luminosity that can be achieved by an accreting black hole when there is hydrostatic equilibrium: $L_{\rm Edd} = 4\pi GM_{\rm BH}cm_{p}/\sigma_{\rm T}$, where $\sigma_{\rm T}$ is the Thomson cross section and $m_{p}$ is the mass of a proton.}. The scaled mass accretion rate of M60, as reported by \citet{Johannsen2012}, is $\dot{m}=1.4\times10^{-8}$ and \citet{Paggi2014} reported a rate of $\dot{m}=7.2\times10^{-9}$. Both of these accretion rates are extremely low. We adopt the arithmetic mean value of these two scaled mass accretion rates, $\dot{m}\sim1.1\times10^{-8}$, for this study. Given the relatively large angular size of $R_{\rm S}$ in M60, this galaxy may provide an excellent opportunity to study the properties of ADAF and jets in LLAGNs with very low mass accretion rates using VLBI observations.


The Very Large Array \citep[VLA,][]{Stanger1986, Temi2022, Grossova2022} and Low-Frequency Array \citep[LOFAR,][]{Capetti2022} images of M60 
exhibit a typical Fanaroff-Riley type I \citep[FR I,][]{Fanaroff1974} radio galaxy morphology on kpc scales,  which includes a compact core and double side jets. These jets are also associated with X-ray cavities \citep{Shurkin2008, Dunn2010, Paggi2014}. The high-resolution VLA A-configuration images \citep{Temi2022, Grossova2022} or high-frequency images \citep{DiMatteo1999} of M60 only display a compact core. Submillimeter Array (SMA) observations detected significant variability at 223~GHz in M60 in March 2016 \citep{Lo2023}. The radio flux density of M60 at 223~GHz varied from not-detection ($\leq8.1$~mJy) on 3 March, increasing to $15.9\pm2.1$~mJy on 5 March, and then decreasing to $6.7\pm1.6$~mJy on 13 March. Infrared observations and dust absorption map suggest that M60 is a cold-gas-free galaxy with no signs of star formation \citep{Temi2022}. Thus, it could be an ideal candidate for VLBI observations to investigate outflow or ADAF radio emission properties. Furthermore, VLBI observations could provide more direct evidence for the presence of an accreting SMBH in M60 \citep[e.g.,][]{Maoz1995, Maoz1998, Genzel2000, Ghez2005, Shen2005} and independently constrain its SMBH mass \citep{Johannsen2010, Johannsen2012}.

This paper is structured as follows. In Section \ref{sec:data}, we introduce the radio data used, which include high-resolution Very Long Baseline Array (VLBA) data and low-resolution archival radio flux density. Section \ref{sec:image} presents the imaging results, and Section \ref{sec:radio} discusses the radio origins of the compact component. Section \ref{sec:bh} provides independent evidence supporting the presence of an SMBH in M60. Finally, Section \ref{sec:conclusion} summarizes the conclusions.

\section{Data reduction and analysis} \label{sec:data}

\subsection{VLBA observations and data reduction}
In the VLBA Calibrator Survey 12 (program code BP252AF, principal investigator: Leonid Petrov), the radio AGN in M60 was observed simultaneously at 4.4 and 7.6 GHz for a duration of 3 min on July 26, 2022. During the experiment, the stations Kitt Peak (\texttt{KP}) and North Liberty (\texttt{NL}) were out, and the station Pie Town (\texttt{PT}) failed to record data. As a result, seven VLBA stations contributed valid data. The experiment was carried out with circular right polarization at a data rate of 4096 Mbps, using 8 subbands (4 at 4.4 GHz and 4 at 7.6 GHz), 128 MHz per subband, and 2-bit quantization. The DiFX2 software correlator \citep{Deller2011} at the National Radio Astronomy Observatory (NRAO) was employed to correlate the data, utilizing a 0.0625-MHz frequency resolution and a 0.1-s integration time. 


The correlation data were calibrated in the Astronomical Image Processing System \citep[\texttt{AIPS} version 31DEC21;][]{Greisen2003}. After the data were imported into \texttt{AIPS}, we corrected the correlation amplitudes due to minor sampling errors, and then ran amplitude calibration based on the system temperature data and antenna gain curves. NRAO recently announced the default antenna gain values of PT and NL at C band (4-8~GHz) between 11 Feb 2020 and 5 April 2023 were incorrect due to incorrect focus. To fix the issue, we used the new gain curve table provided by NRAO \footnote{\href{https://science.nrao.edu/facilities/vlba/data-processing/vlba-7ghz-flux-density-scale}{https://science.nrao.edu/facilities/vlba/data-processing/vlba-7ghz-flux-density-scale}}. To align phase across sub-bands, we performed fringe fitting on the 1-min data from the $\sim$1-Jy bright calibrator B1116$+$128. After the manual phase calibration, we merged the subbands in halves (i.e., the 4.4 and 7.6~GHz data are combined, respectively) to enhance the Signal Noise Ratio (SNR) of the solution and conducted global fringe fitting on the data of M60 with 0.75~min solution interval. All the solutions had SNRs greater than 4. Finally, we applied all the solutions to the data,  averaged all the frequency channels in each subband, and split the data at 4.4 and 7.6 GHz into two files.

\begin{figure*}[ht!]
\includegraphics[width=0.5\textwidth]{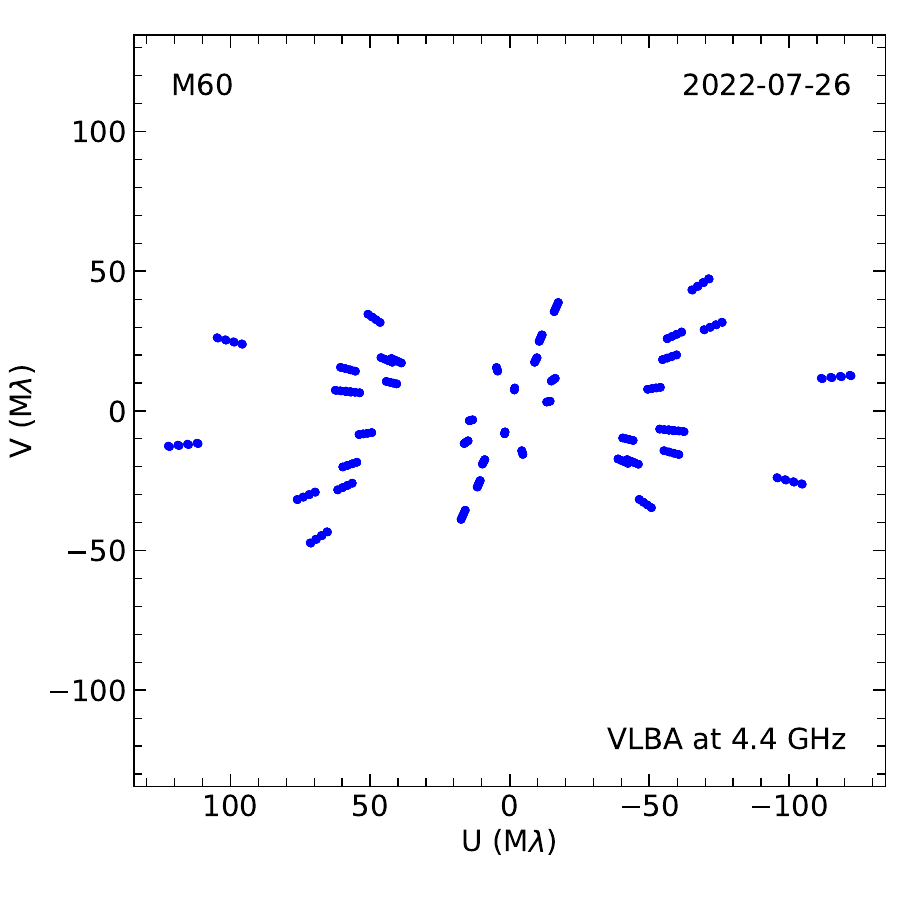}
\includegraphics[width=0.5\textwidth]{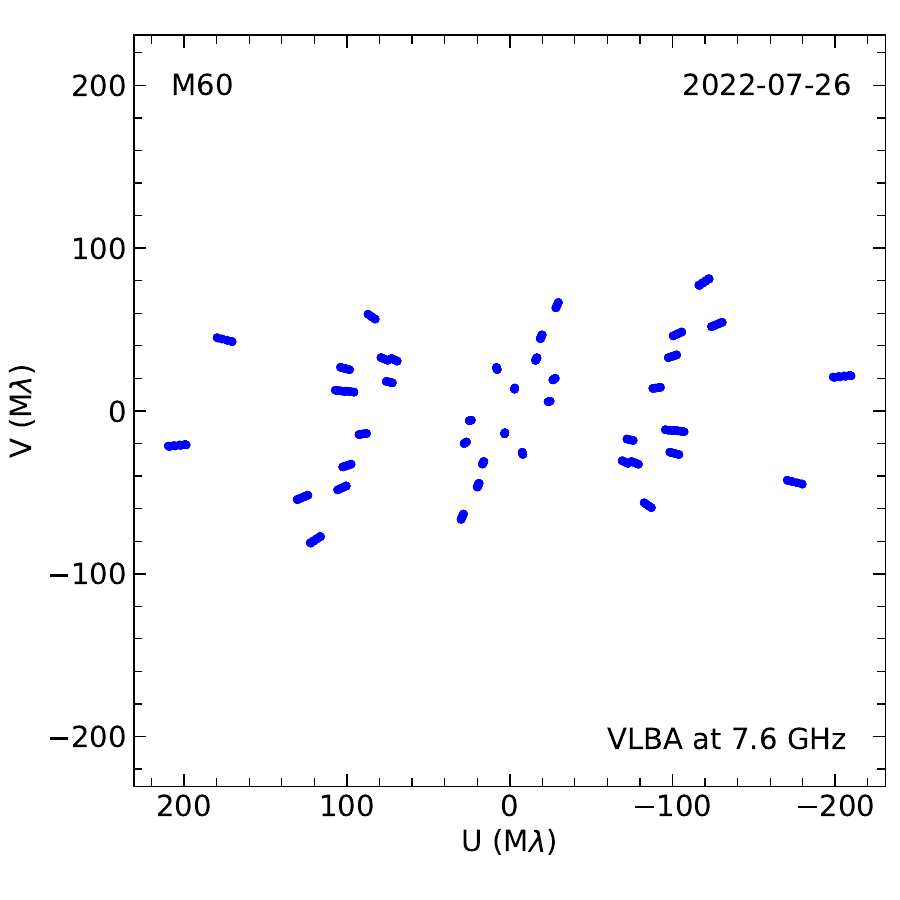}\\
\caption{ $(u,v)$ coverage for M60 at 4.4 GHz (left) and 7.6 GHz (right). The $(u,v)$ coordinates for each pair of antennas are the baseline length projected from the source in units of the observing wavelength $\lambda$ and are given for conjugate pairs. \label{fig:uvcov}}
\end{figure*}

The $(u,v)$ coverage of M60 is shown in Figure \ref{fig:uvcov}. The 4.4 and 7.6 GHz data have similar characteristics, with a sparse $(u,v)$ coverage. The maximum side lobe was approximately 72\% of the primary lobe.

The radio source M60 is relatively weak (correlation amplitude $\sim$20~mJy). Before running the amplitude self-calibration on the M60 data, we also did a test to verify the reliability of the step. In the same geodetic VLBI experiment, the compact calibrator B1241$+$166 with a total flux density of $\sim$380~mJy at 4.4 GHz was also observed. Because both M60 and B1241$+$166 have a small separation of $4.8^\circ$ and were observed in two neighboring scans, their data were supposed to suffer fairly consistent amplitude errors. We did deconvolution and self-calibration on the data of the pair of sources in \texttt{Difmap}. The amplitude self-calibration solutions looked quite similar and were fully as expected. Therefore, it is safe to run the self-calibration of amplitude on the data of M60.  

In view of the positive results above, we transferred the amplitude self-calibration solutions from the bright calibrator B1241$+$166 to the nearby target M60 in \texttt{AIPS}. The gain correction factors of B1241$+$166, derived from \texttt{gscale} command in \texttt{Difmap} \citep{Shepherd1997}, are reported in Appendix~\ref{sec:gain} (Table~\ref{tab:gain}). We noticed that the gain correction factors of the station LA were abnormally larger than those of other stations. After contacting VLBA support scientists, we figured out that these large correction factors resulted from improper antenna gain measurements.

\begin{figure*}[ht!]
\includegraphics[width=0.5\textwidth]{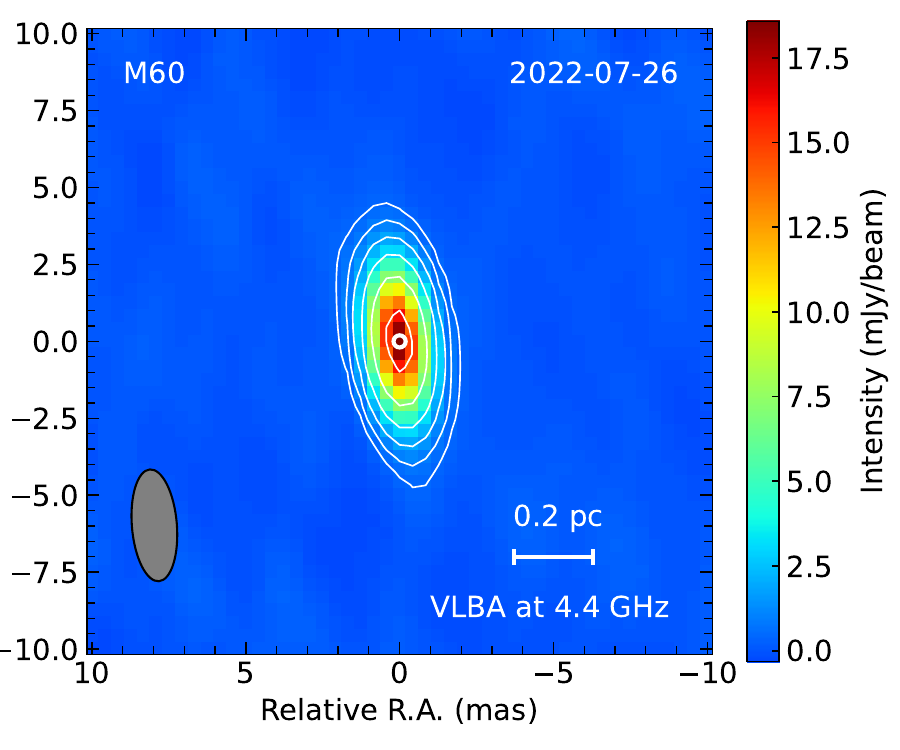}
\includegraphics[width=0.5\textwidth]{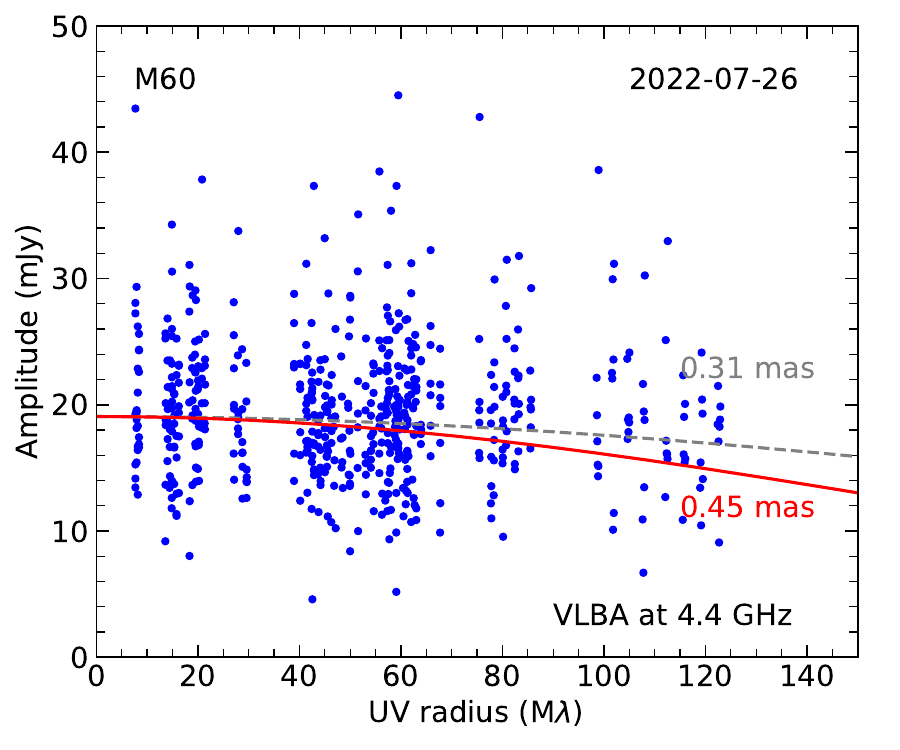}\\
\includegraphics[width=0.5\textwidth]{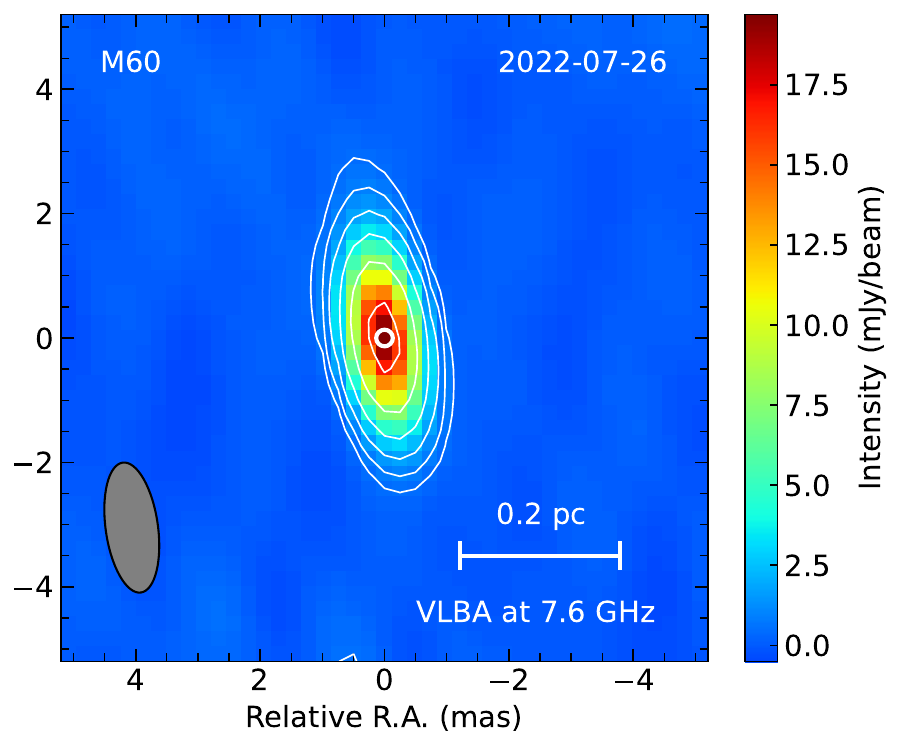}
\includegraphics[width=0.5\textwidth]{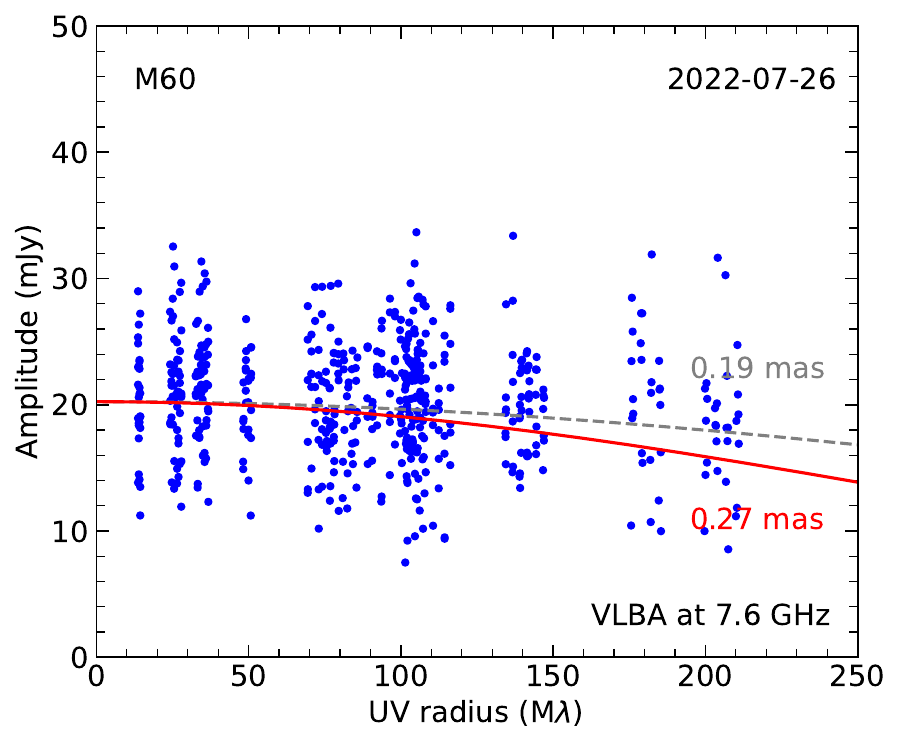}
\caption{ 
The VLBA \textsc{clean} images (Left) and observed visibility (Right) of M60 (NGC 4649). The upper panels are 4.4~GHz results, while the bottom panels are 7.6~GHz results. The peak brightness of 4.4/7.6~GHz image is $18.83/17.40\,{\rm mJy\,beam^{-1}}$. The rms of the 4.4/7.6~GHz image is $0.16/0.17~\rm{mJy\,beam^{-1}}$.  The contour levels of 4.4/7.6~GHz image are $0.48/0.51\,{\rm mJy\,beam^{-1}}\times(-1, 1, 2, 4,8, 16)$ and the color bar is in $\rm mJy\,beam^{-1}$.  The beam FWHM of 4.4~GHz image is $3.68\times1.50\,{\rm mas}$ at $6\fdg11$ and the beam FWHM of 7.6~GHz image is $2.12\times0.86\,{\rm mas}$ at $7\fdg28$. The thick white lines at the image center represent the upper limit (99.7\% confidence) of the de-convoluted angular size. The correlation amplitude versus the radius in the ($u$, $v$) plane.  The visibility data were calibrated. The gray dashed line and red solid line represent the models using the angular sizes at 50\% and 99.7\% confidence levels. 
\label{fig:image}}
\end{figure*}

After calibrating out the large amplitude errors with the nearby bright calibrator B1241$+$166 in \texttt{AIPS}, we did deconvolution and self-calibration in \texttt{Difmap}.  Initially, we fitted the circular Gaussian model to the visibility data and then performed phase self-calibration with a solution interval of 1.5~min. After a few iterations, we started to run the amplitude self-calibration. The amplitude correction factors of M60 were also listed in Appendix~\ref{sec:gain} (Table~\ref{tab:gain}). All the correction factors are quite close to 1, in particular at 4.4 GHz. The \texttt{selfcal} command was then run with a solution interval of 3~min, allowing the amplitude and phase to change simultaneously. The model fitting and self-calibration processes were repeated until the reduced chi-square $\chi_{\nu}^{2}$ was minimized. To obtain accurate uncertainty estimates, we scaled the data weights to set $\chi_{\nu}^{2} = 1$. The data at 4.4 and 7.6 GHz were reduced in the same method. We also tried the elliptical Gaussian model. Because some best-fitting parameters, in particular the position angle of the major axis, had large uncertainties, the more complex model was not considered in the paper. The final clean images of M60 at 4.4 and 7.6~GHz are presented in Figure~\ref{fig:image}.

To accurately characterize the uncertainties of the parameters of the Gaussian model, in particular the angular size, we also used the Markov Chain Monte Carlo (MCMC) method to fit the circular Gaussian mode to the visibility data calibrated in \texttt{Difmap}. The used MCMC program \texttt{MCMCUVFit} was developed by us. More details of the program are available in Appendix \ref{sec:mcmc}. The results of the MCMC method are presented in Table \ref{tab:model}. The calibrated visibility and the model from the MCMC output are shown in the right panels of Figure \ref{fig:image}. 

We also tried the MCMC and least-square fitting methods to analyze the M60 visibility data using phase-only self-calibration in \texttt{Difmap}. The best-fitting parameters show slightly larger uncertainties but are fully consistent with the above results using the phase and amplitude self-calibration.


\subsection{Non-simultaneous radio spectrum}
To analyze the origin of the radio emission in M60, we collected radio flux density measurements in some online data archives and literature. To exclude possible diffuse emission from the relic jet, we used the measurements from the VLBA data and the VLA data observed with A-configuration at $\leq15$~GHz. At higher frequencies, the contribution of the extended steep spectrum relic jet is quite low and can be ignored. Therefore, we included low-resolution data. The VLBA data are the first VLBI detected flux density that is presented in Appendix \ref{sec:mcmc}. Most data extracted from the NRAO VLA Archive Survey \citep[NVAS,][]{Crossley2008} \footnote{\href{https://www.vla.nrao.edu/astro/nvas/}{https://www.vla.nrao.edu/astro/nvas/}}. M60 was observed by ALMA at $\sim$350~GHz to study its black hole mass, central parsec gas dynamics, and event horizon detectability by CO lines (program code 2016.1.01135.S and 2017.1.00830.S, Principal Investigator: Neil Nagar). \citet{Temi2022} reported non-detection of the CO lines. We reduced these ALMA data to extract flux densities from the radio continuum spectrum. The data reduction was carried out in Common Astronomy Software Applications (CASA) casa-release-5.1.1 \citep{McMullin2007} with ALMA Pipeline-Cycle5-R2-B \citep{ALMAPipeline}. The resulting radio flux density of M60 at 350~GHz is $1.7\pm0.2$~mJy on 18 July 2017 and $1.1\pm0.1$~mJy on 24 December 2017. We also include four published data points: one data point was observed by VLA with A-configuration at 1.5 \citep{Temi2022, Grossova2022}, two data points were observed by VLA with D-configuration at 22 and 43~GHz respectively \citep{DiMatteo1999}, one data point was observed by submillimetre common-user bolometer array (SCUBA) at 150~GHz \citep{DiMatteo1999}. These error bars include systematic errors: 5 percent for the VLBA results and 3 percent for the VLA results. All these flux density measurements are presented in Table \ref{tab:flux}. There is no hint of strong variability from these non-simultaneous observations, which span approximately four decades, except the SMA data, which is not used.

\section{Results and discussion} \label{sec:diss}

\subsection{Radio images of M60} \label{sec:image}

Recently, the Astrogeo Center has precisely determined the location of the radio source in M60. According to the Radio Fundamental Catalog (RFC)\footnote{\href{http://astrogeo.org/rfc/}{http://astrogeo.org/rfc/}}, which is available to the public, the coordinates are RA=12:43:39.9715 and Dec=+11:33:09.688 (with an error of 1.8 mas). In Gaia Data Release 3 (DR3), the optical coordinates of M60 are given as RA=12:43:39.9708 (with an error of 6.2 mas) and Dec=+11:33:09.6882 \citep[with an error of 2.8 mas;][]{Gaia2022}. The RFC position of M60 is located east of the Gaia DR3 position, with a separation of 9.9 mas. This indicates that the radio position is consistent with the optical position.


The left panels of Figure \ref{fig:image} display the total intensity images of M60 at 4.4 and 7.6~GHz. The peak brightness of the 4.4/7.6~GHz image is $18.58/19.69\,{\rm mJy\,beam^{-1}}$ and the image rms is 0.16/0.17~mJy\,beam$^{-1}$, respectively. The image scale is $0.08~{\rm pc\,mas^{-1}}$ and the average beam size at 4.4/7.6~GHz is 2.31/1.33~mas. The total flux density at 4.4 and 7.6 GHz is $19.1\pm1.0$~mJy and $20.2\pm1.0$~mJy, respectively, with the error mainly due to the systematic error ($5\%$). The radio component remains unresolved, with an upper limit size of $\leq 0.45$~mas (0.036~pc, or 83~$R_{\rm S}$) at 4.4~GHz, and $\leq$0.27~mas (0.022~pc, or 50~$R_{\rm S}$) at 7.6~GHz at a confidence level of $99.7\%$. These sizes are roughly a fifth of the restore beam. The fitted component sizes at both bands are on centi-pc-scale. The relatively low image sensitivity did not allow us to detect the emission beyond the compact core, and the sub-pc-scale jet might be resolved out or substantially weaker than the image sensitivity can detect.

The VLBA data of M60 suggest a spectral index of $\alpha_{\rm 4.4\,GHz}^{\rm 7.6\,GHz}=0.11\pm0.13$ ($S_{\nu}\propto \nu^{+\alpha}$), implying a slightly inverted spectrum at 4.4-7.6~GHz. Figure~\ref{fig:spix} displays the non-simultaneous radio spectrum of M60, which is marginally inverted at $\leq21.6$~GHz with $\alpha_{\rm thick}=0.12\pm0.04$  and steep at $>21.6$~GHz with $\alpha_{\rm thin}=-1.5\pm0.1$ (see Section \ref{sec:jet}). The brightness temperature of the unresolved component is $T_{\rm B}(4.4{\rm \,GHz})\geq6.0\times10^{9}$~K and $T_{\rm B}(7.6{\rm \,GHz})\geq5.9\times10^{9}$~K, which is much higher than the typical brightness temperature of thermal radiation ($\leq10^5$~K).

The morphology of M60 on both the kpc and centi-pc scales, its slightly inverted radio spectrum, and its extremely high brightness temperature all point to the radio emission being caused by an AGN rather than star formation. This is further supported by the lack of cold gas, which is necessary for star formation, as evidenced by the non-detection of [C II] emitting gas with SOFIA, the non-detection of CO line by ALMA, and the absence of dust in the HST dust absorption map \citep{Temi2022}.

\subsection{Origin of the centi-pc-scale compact component} \label{sec:radio}

In the vicinity of LLAGN SMBH, radio emission could be generated by outflows such as jets \citep{Blandford1977, Falcke1999, Chen2021} and winds \citep{Blandford1982, Blandford1999}, or an ADAF \citep{Narayan1994, Mahadevan1997, Manmoto1997}. However, in the case of M60, no extended jets or winds exceeding $0.85~\mathrm{mJy\,beam^{-1}}$ were detected in the VLBI images. Therefore, we only consider two possible origins: the jet base and the ADAF. Both of these origins have a peaked continuum spectrum in the radio bands. The jet base has an inverted spectrum at low frequency with a spectral index of $0<\alpha\leq2.5$, and a steep spectrum at high frequency \citep{Falcke1999}. On the other hand, the ADAF has a similar continuum spectrum but with a spectral index of $\alpha\sim0.4$ at low frequency \citep{Mahadevan1997}. A comprehensive analysis of these possibilities is provided in the following sections.

\subsubsection{\textbf{Jet base}} \label{sec:jet}

\begin{figure*}[ht!]
\includegraphics[width=0.495\textwidth]{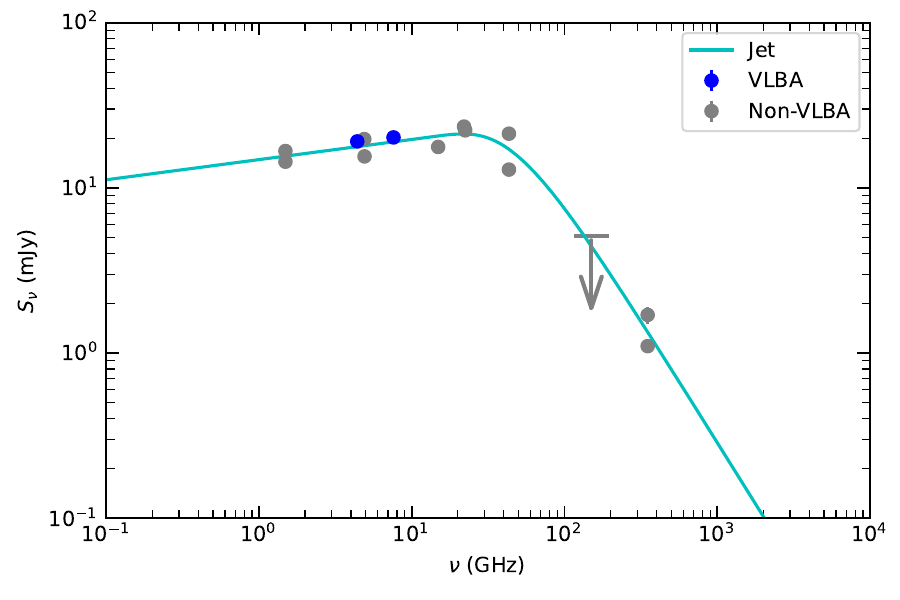}
\includegraphics[width=0.495\textwidth]{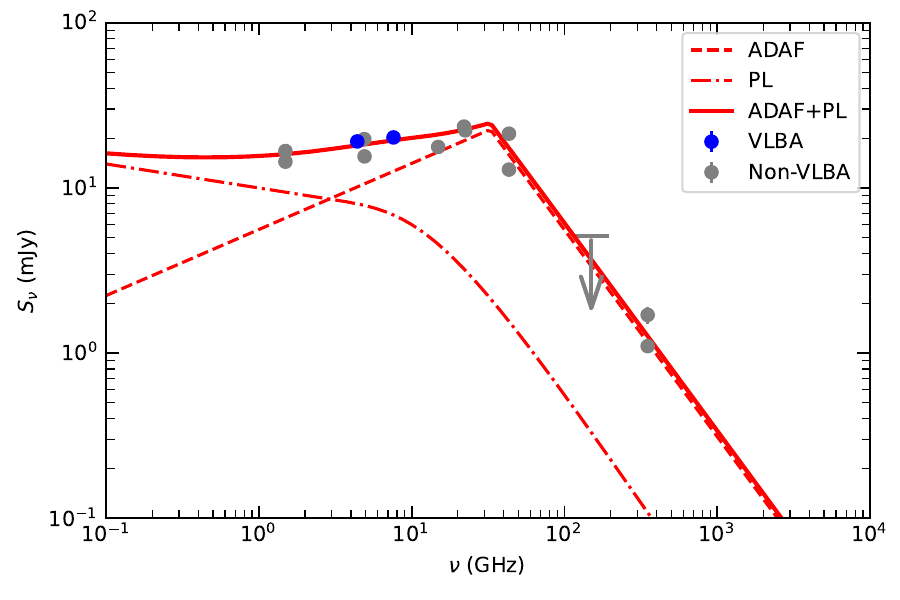}\\
\centering
\includegraphics[width=0.5\textwidth]{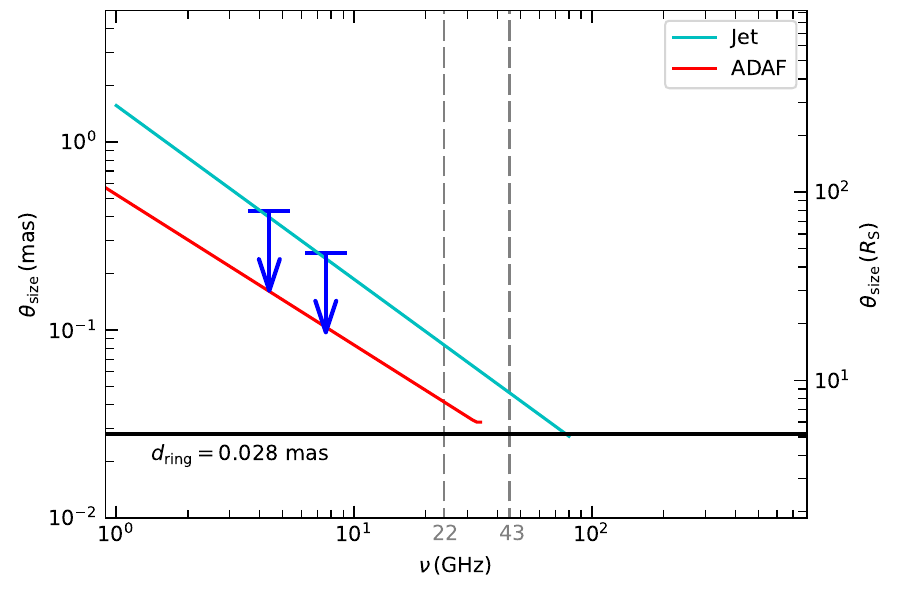}
\caption{The non-simultaneous radio spectrum and the observed angular sizes of M60. Top Left: The gray and blue-filled circles represent the Non-VLBA and VLBA data, respectively. The cyan solid line depicts the best-fitting results from a peaked spectrum of jet base emission. Top Right: The radio spectrum is fitted to a model composed of an ADAF component and a power-law component. The dashed and dashed-dotted red lines represent the contribution from the ADAF and power-law components, respectively. The total flux density is shown by the red solid line. Bottom: The blue arrows indicate the upper limit of the observed angular size. The horizon black solid line predicts the photon ring diameter based on the dynamics mass of the SMBH. The grey dashed vertical line indicates the available VLBI observing frequencies close to the peak frequency. The cyan solid lines represent core size predicted by a jet model with characteristic electron Lorentz factor of $\gamma_{e,0}=230$. The core size predicted by the ADAF model, which only has thermal electrons, is shown by the red solid line.
\label{fig:spix}}
\end{figure*}

The radio continuum spectrum of M60 is shown in Figure \ref{fig:spix}. To measure the variability of M60, we employ the variability index $m_{\rm var}=\sigma/<S>$, where $<S>$ and $\sigma$ are the mean and standard deviation of the flux density, respectively. The $m_{\rm var}$ at 1.5~GHz and 4.9~GHz is approximately 7.4\% and 11.9\%, respectively, indicating a low variability. The radio flux density at 223~GHz detected by SMA showed significant variability during March 2016 \citep{Lo2023}, but was not used in the continuum spectrum fitting. The M60 radio spectrum is characterized by a peaked shape, which can be attributed to the synchrotron radiation of jets \citep{Konigl1981, Chen2017}. Inverted and steep spectral indexes indicate the optically thick and thin parts, respectively. To fit the radio spectrum of M60, we use a smooth broken power law model proposed by \citet{Callingham2017}.
The flux density $S_{\nu}$ is given by
\begin{equation}\label{eq:ps}
    S_{\nu} = \frac{S_{\rm t}}{1-e^{-1}} \left(1-e^{-(\nu/\nu_{\rm t})^{\alpha_{\rm thin}-\alpha_{\rm thick}}}\right) \left(\frac{\nu}{\nu_{\rm t}}\right)^{\alpha_{\rm thick}}
\end{equation}
where $\alpha_{\rm thick}$ and $\alpha_{\rm thin}$ are the spectral indices in the optically thick and optically thin regions, respectively, and $S_{\rm t}$ is the flux density at the turnover frequency $\nu_{\rm t}$. The data used to fit is presented in Table \ref{tab:flux}, and a variable error 10\% ($\sigma_{\rm var}=10\%\times S_{\rm tot}$) was added to the flux density error, i.e., $\sigma=\sqrt{\sigma_{\rm m}^{2}+\sigma_{\rm var}^{2}}$ where $\sigma_{\rm m}$ is the measurement error. The fitting results are $\nu_{\rm t} = 51_{-7}^{+8}$~GHz, $S_{\rm t} = 15_{-1}^{+1}$~mJy, $\alpha_{\rm thick} = 0.12_{-0.04}^{+0.04}$ and $\alpha_{\rm thin} = -1.5_{-0.1}^{+0.1}$. The flux density of the model is plotted with a solid cyan line in Figure \ref{fig:spix} (top left). It is important to note that $\nu_{\rm t}$ is not the peak frequency but the turnover frequency. The peak frequency $\nu_{\rm p}$ and peak flux density $S_{\rm p}$ can be numerically derived from model parameters. By substituting the parameters of the fit into equation \ref{eq:ps}, we obtain $\nu_{\rm p}\sim21.6~$GHz and $S_{\rm p}\sim21.2~$mJy.

Equation 8 in \citet{Falcke1999} can be used to determine the jet power $Q_{\rm jet}$ and the inclination angle $i$ of the jet in relation to the line of sight. The characteristic electron Lorentz factor of the jet, denoted as $\gamma_{\rm e,0}$, is a free parameter. We set it to $\gamma_{\rm e,0}=230$, which is lower than the median value of 300, to ensure that the simulated jet size is in line with the core size-frequency relation (Figure \ref{fig:spix}, bottom). This yields an estimated jet power of $Q_{\rm jet} \sim 5.7\times10^{41}~{\rm erg\,s^{-1}}$ and an inclination angle of $i\sim30^\circ$. 


\subsubsection{ADAF}\label{sec:adaf}

The low luminosity \citep{DiMatteo1997, DiMatteo1999, Paggi2014} and low accretion rate \citep{Johannsen2012, Paggi2014} of M60 suggest that its accretion disk is an ADAF, which exhibits radio emission in the inner disk region. This ADAF model has a peaked spectrum with an inverted spectral index of approximately 0.4 before the peak frequency, which is due to synchrotron cooling, and a steep spectral index after the peak frequency, which is caused by Compton cooling \citep{Manmoto1997, Mahadevan1997}. \citet{Mahadevan1997} provided an analytical expression for the ADAF radiation spectrum by a self-similar solution, which can accurately describe the main features of ADAF and can be used to fit the observed data. In this model, the radio emission is generated by synchrotron radiation from thermal electrons.


The parameters used to characterize the ADAF model include the scaled black hole mass $m$, the scaled accretion rate $\dot{m}$, the viscosity parameter $\alpha$, the ratio of gas pressure to total pressure $\beta$, and the fraction of viscous energy transferred to electrons $\delta$. The black hole mass \citep[$m=4.5\times10^{9}$,][]{Shen2010} and accretion rate \citep[$\dot{m}\sim1.1\times10^{-8}$, averaged from two independent measurements,][]{Johannsen2012, Paggi2014} were determined by observations and kept constant. The other three parameters $\alpha$, $\beta$, and $\delta$ were allowed to vary in order to fit the radio continuum spectrum of M60. We tried to fit the data with the \citet{Mahadevan1997} model and found that the low-frequency band could not be accurately modeled because the observed spectral index $\alpha_{\rm thick}\sim0.12$ was not consistent with the ADAF model. Therefore, a simple ADAF model with only thermal electrons cannot fully explain the observed radio spectrum of M60.

We included a broken power law (BPL) component in the model to refine the fit of the low-frequency data. This BPL component has the same expression as equation \ref{eq:ps}, but allows for a negative value for $\alpha_{\rm thick}$, that is, a flat spectrum. The fitting results yielded the following parameters: $\alpha=2.4_{-1.1}^{+1.7}\times10^{-4}$, $\beta=0.96_{-0.03}^{+0.01}$, $\delta=0.61_{-0.30}^{+0.25}$, $\nu_{\rm t}^{\rm BPL}=17.4_{-7.9}^{+8.9}$~GHz, $S_{\rm t}^{\rm BPL}=4.3_{-1.2}^{+1.2}$~mJy, $\alpha_{\rm thick}^{\rm BPL}=-0.13_{-0.13}^{+0.09}$, $\alpha_{\rm thin}^{\rm BPL}=-1.3_{-0.4}^{+0.2}$. The fitting results are shown in Figure \ref{fig:spix} (top right).

The BPL component can be attributed to the synchrotron radiation of nonthermal electrons in the ADAF \citep{Yuan2003, Ozel2000, Bandyopadhyay2019, Cho2022}. This further indicates that the core size at lower frequencies is larger than the pure thermal electron model. Moreover, the core size versus frequency relation is more complicated, and no straightforward solution can be derived. Therefore, we have only plotted the core size versus frequency relation for the pure thermal electron model, considering it as a lower bound (Figure \ref{fig:spix} bottom).


\subsection{New evidence for the existence of a SMBH in M60}\label{sec:bh}

The stellar dynamics \citep{Shen2010} and X-ray hot gas hydrodynamics \citep{Humphrey2008, Paggi2014, Paggi2017} of M60 suggest the presence of a massive dark object at its center. This could be an SMBH or a cluster of brown dwarfs or stellar remnants \citep{Maoz1995, Maoz1998}. Here, we present a new argument in favor of the central massive object being an SMBH. Previous studies have demonstrated that if the maximum lifetime of a putative dark cluster is much less than the age of the host galaxy, due to evaporation and physical collision, it could not survive to present \citep{Maoz1995, Maoz1998, Genzel2000, Ghez2005, Shen2005}. With a black hole mass of $4.5\times10^{9}\,M_{\odot}$ and a core size at 7.6 GHz of $0.27\,{\rm mas}$, the half mass density of the central object is $\rho_{\rm h} \geq 5.4\times10^{13}\,M_{\rm \odot}{\rm \,pc^{-3}}$. If the central object is a dark cluster, its lifetime would be $\tau \leq 8.4\times10^{9}\,{\rm yr}$, which is less than the lifetime of the host galaxy, $\sim10^{10}\,{\rm yr}$ (see Figure \ref{fig:tau} and the detailed calculation method presented in Appendix \ref{sec:lifetime}). Therefore, it is more likely that the centi-pc-scale radio core in M60 is associated with an SMBH.

Johannsen et al. (2010) proposed a method to estimate the mass of SMBH by imaging its compact radio structure with VLBI. They showed that the photons emitted around the SMBH form a ring-like structure, its diameter is determined by the mass of the SMBH ($d_{\rm ring}=5.2\, R_{\rm S}$) and are not significantly affected by the spin of the SMBH, the angle of inclination, or the observation frequency. Thus, the mass of the SMBH can be estimated by the size of the associated compact radio source. Taking the core size of M60 at 7.6 GHz ($\theta_{\rm size}\leq0.27\,{\rm mas}$) as the upper limit of the photon ring diameter, an upper limit for the SMBH mass can be derived: $M_{\rm BH} \leq 4.3\times10^{10}\, M_{\rm \odot}$. This independent derived upper limit for the black hole mass is consistent with the mass determined by stellar dynamics ($(4.5\pm1.0)\times10^{9}~M_{\odot}$, Shen et al. 2010) and the mass estimated from X-ray hot gas hydrodynamics ($(5.1\pm0.9)\times10^{9}~M_{\odot}$, Humphrey et al. 2008, Paggi et al. 2014, 2017).

\section{Conclusions} \label{sec:conclusion}

We presented high-resolution VLBA observations conducted in snapshot mode at 4.4 and 7.6~GHz, alongside radio flux density data that include low-resolution data. The resulting VLBA image revealed compact components with a flux density of $19.1\pm1.0$~mJy at 4.4~GHz and $20.2\pm1.0$~mJy at 7.6~GHz. The size of the components, estimated by MCMC fitting, was $\leq83 R_{\rm S}=0.036\,\mathrm{pc}=0.45\,$mas at 4.4~GHz and $\leq50 R_{\rm S}=0.022\,\mathrm{pc}=0.27~$mas at 7.6~GHz. The brightness temperature of the compact components in both bands was $T_{\rm B}\geq6\times10^{9}$~K. The multiband radio flux density of M60 displayed a peaked spectrum. These results enable us to confidently identify the compact component as the radio core of M60. The radio core detected by VLBA within M60 could be due to synchrotron radiation of nonthermal electrons in the jet base, or it could also be attributed to synchrotron radiation generated by a combination of thermal and nonthermal electrons present within ADAF. The high mass density ($\rho_{\rm h} \geq 5.4\times10^{13}\,{\rm M_{\odot}\,pc^{-3}}$) suggests that the central object is an SMBH rather than a dark cluster. The black hole mass was restricted to $M_{\rm BH}\leq 4.3\times10^{10}\,M_{\odot}$, which is in agreement with the dynamical mass.

Our findings indicate that M60 could be a promising object for further deep VLBI observations at $\geq$15~GHz. This would enable us to accurately determine the diameter of the potential photon ring and the mass of its SMBH, as well as to investigate the jet base and ADAF in M60. We have been granted permission for 16-Gbps observations on M60 at 22 and 43~GHz using the Korean VLBI Network (KVN) and some stations in the European VLBI Network (EVN). This project is scheduled to be observed in the near future.


\begin{acknowledgments}

This work is supported by the National Natural Science Foundation of China (No.12203014). X.-P. Cheng was supported by Brain Pool Program through the National Research Foundation of Korea (NRF) funded by the Ministry of Science and ICT (2019H1D3A1A01102564). We thank Feng Wang, Ying Mei, and Hui Deng for their insightful discussions during the initial stages of our work. This research has made use of the NASA/IPAC Extragalactic Database (NED), which is operated by the Jet Propulsion Laboratory, California Institute of Technology, under contract with the National Aeronautics and Space Administration. The National Radio Astronomy Observatory is a facility of the National Science Foundation operated under cooperative agreement by Associated Universities, Inc. This paper makes use of the following ALMA data: ADS/JAO.ALMA\#2016.1.01135.S and ADS/JAO.ALMA\#2017.1.00830.S. ALMA is a partnership of ESO (representing its member states), NSF (USA) and NINS (Japan), together with NRC (Canada), MOST and ASIAA (Taiwan), and KASI (Republic of Korea), in cooperation with the Republic of Chile. The Joint ALMA Observatory is operated by ESO, AUI/NRAO and NAOJ.

\end{acknowledgments}
%

\vspace{5mm}
\facilities{VLBA (NRAO), VLA (NRAO), ALMA.}


\software{AIPS \citep{Greisen2003},
          CASA \citep{McMullin2007},
          Difmap \citep{Shepherd1997},  
          DiFX \citep{Deller2011}, 
          MCMCUVFit,
          Astropy \citep{Astropy2018}.
          }

\begin{deluxetable*}{cccccccc}
\tablecaption{VLBI results of M60 at C band} \label{tab:model}
\tablehead{
\colhead{$\nu$}& \colhead{Beam} & \colhead{$S_{\rm p}$} & \colhead{rms}  & \colhead{$S_{\rm tot}$}  & \colhead{$\theta_{\rm core}$} & \colhead{ $x$} & \colhead{$y$}\\
\colhead{(GHz)}& \colhead{(mas$\times$mas,\,$^\circ$)} & \colhead{(${\rm mJy\,beam^{-1}}$)}  & \colhead{($\mathbf{\rm mJy\,beam^{-1}}$)}  & \colhead{(mJy)} & \colhead{(mas/$R_{\rm S}$)} & \colhead{(mas)} & \colhead{(mas)} 
}
\colnumbers
\startdata
4.4 & 3.65$\times$1.46,\,4.69 & 18.58 & 0.16 & $19.1\pm1.0$ &  $\leq0.45/\leq83$ & $0.008_{-0.006}^{+0.006}$ & $0.015_{-0.015}^{+0.015}$ \\
7.6 & 2.11$\times$0.84,\,7.99  & 19.69 & 0.17 & $20.2\pm1.0$ & $\leq 0.27/\leq50$ & $0.007_{-0.004}^{+0.004}$ & $0.008_{-0.009}^{+0.009}$ \\ 
\enddata
 \tablecomments{Column 1: Observation frequency in GHz; Column 2:  Restore bean shape (Major axis and Minor axis in mas, Position angle in degree); Column 3: Peak flux density in ${\rm mJy\,beam^{-1}}$; Column 4: Image rms in $\rm mJy\,beam^{-1}$; Column 5: Total flux density in mJy; Columns 6: Core size in mas or $R_{\rm S}$ with 99.7\% confidence; Column 7 \& 8: Position relative to image center in mas.}
\end{deluxetable*}

\begin{deluxetable*}{ccccc}
\tablecaption{Radio flux density for continuum spectrum fitting} \label{tab:flux}
\tablehead{
 \colhead{Telescope} & \colhead{Epoch}  & \colhead{$\nu$} &  \colhead{$S_{\nu}$}  & \colhead{Reference} \\
\colhead{}  & \colhead{(Y-M-D)} & \colhead{(GHz)} & \colhead{(mJy)} & \colhead{} 
}
\colnumbers
\startdata
 VLA/A & 1982-04-04 & 1.5 & 16.7$\pm$0.5 & NVAS \\
 VLA/A & 1982-04-04 & 4.9 & 15.5$\pm$0.5 & NVAS \\
 VLA/A & 1982-04-04 & 15 & 17.7$\pm$0.5 & NVAS \\
 VLA/A & 1984-12-15 & 4.9 & 19.7$\pm$0.6 & NVAS \\ 
 VLA/A & 1991-08-25 & 1.5 & 14.4$\pm$0.4 & T22, G22 \\ 
 VLA/D & 1997-12-07 & 22 & 23.5$\pm$0.7 & D99 \\
 VLA/D & 1997-12-07 & 43 & 12.9$\pm$0.4 & D99 \\
 SCUBA & 1998-01-30 & 150 & $\leq$5.1 & D99 \\
 VLA/C & 1998-12-30 & 22 & 22.3$\pm$0.7 & NVAS \\
 VLA/C & 1998-12-30 & 43 & 21.3$\pm$0.6 & NVAS \\
 ALMA & 2017-07-18 & 350 & 1.7$\pm$0.2 & This work \\
 ALMA & 2017-12-24 & 350 & 1.1$\pm$0.1 & This work \\
 VLBA & 2022-07-26 & 4.4 & 19.1$\pm$1.0 & This work\\
 VLBA & 2022-07-26 & 7.6 & 20.2$\pm$1.0 & This work\\
\enddata
 \tablecomments{Column 1: Telescope; Column 2: Epoch; column 3: Observation frequency in GHz; Column 4:  flux density in mJy; Column 5: Reference. NVAS: \citealp{Crossley2008}, T22: \citealp{Temi2022}, G22: \citealp{Grossova2022}, D99: \citealp{DiMatteo1999}.}
\end{deluxetable*}

\begin{figure*}[ht!]
\centering
\includegraphics[width=0.7\textwidth]{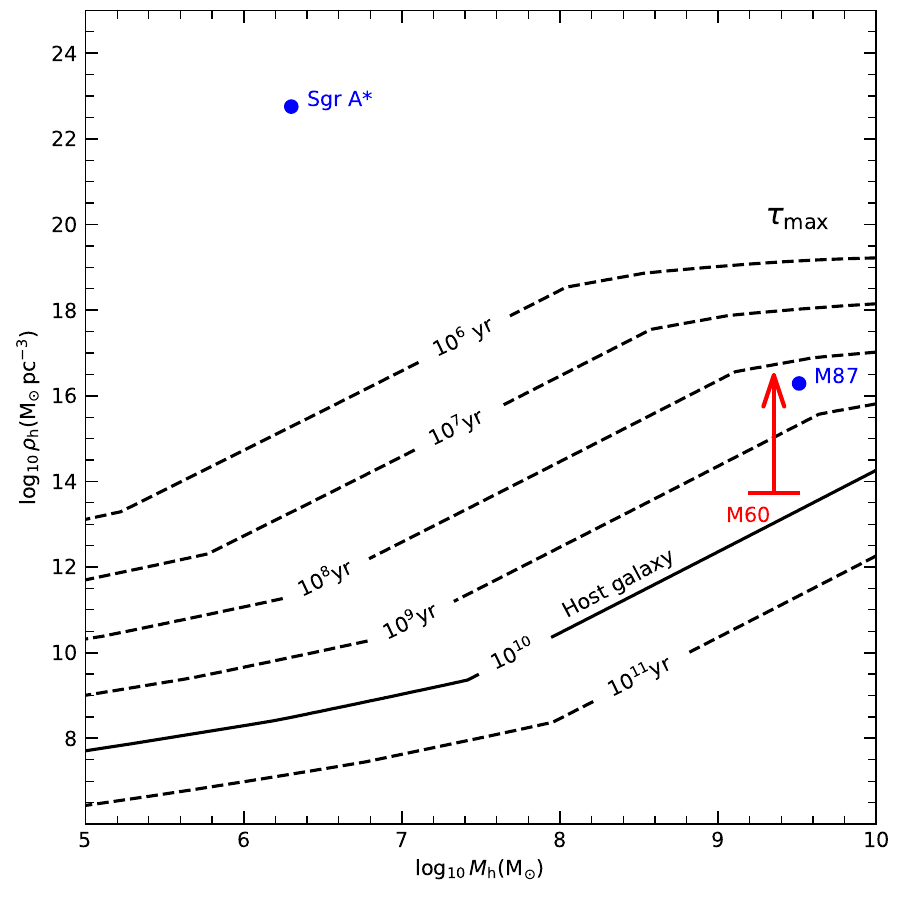}
\caption{The maximum potential lifespan of a dark cluster with half-mass $M_{\rm h}$ and half-mass density $\rho_{\rm h}$. The red data point represents M60 from this work, while the blue data points denote M87 \citep{EHTC2019} and Sgr A* \citep{EHTC2022}. The lifespan of the hypothetical dark cluster in M60 is less than 10 Gyr, suggesting that the central object is a supermassive black hole. \label{fig:tau}}
\end{figure*}

\appendix 

\section{Gain correction factors for amplitude self-calibration}\label{sec:gain}

\begin{deluxetable*}{cccccccccccccccc}
\tablecaption{Gain correction factors of amplitude calibrator 1241+166 and target source M60} \label{tab:gain}
\tablehead{
\colhead{IF} & \colhead{BR} & \colhead{FD}  & \colhead{HN} &  \colhead{LA}  & \colhead{MK} & \colhead{OV} & \colhead{SC} & \colhead{IF}  & \colhead{BR} & \colhead{FD}  & \colhead{HN} &  \colhead{LA}  & \colhead{MK} & \colhead{OV} & \colhead{SC}
}
\startdata
\multicolumn{8}{c}{1241+166 at 4.4~GHz} & \multicolumn{8}{c}{M60 at 4.4~GHz}\\
1 & 1.04 & 1.03 & 0.94 & 1.30 & 0.88 & 0.88 & 1.00 & 1 & 0.96 & 0.97 & 1.01 & 1.02 & 1.01 & 1.04 & 1.04 \\
2 & 0.85 &  0.98 & 0.96 & 1.25 & 0.92 & 0.97 & 1.03 & 2 & 1.01 & 0.97 & 0.99 & 0.98 & 0.98 & 1.03 & 1.01 \\
3 & 0.92 & 0.88 & 0.94 & 1.22 & 0.84 & 0.93 & 1.09 & 3 & 0.98 & 0.95 & 1.02 & 1.01 & 0.98 & 1.02 & 0.98 \\
4 & 1.00 & 0.86 & 0.86 & 1.11 & 0.79 & 0.93 & 0.99 & 4 & 1.03 & 1.02 & 0.95 & 1.05 & 0.97 & 1.03 & 0.90 \\
\hline
\multicolumn{8}{c}{1241+166 at 7.6~GHz} & \multicolumn{8}{c}{M60 at 7.6~GHz}\\
1 & 1.04 & 0.87 & 1.03 & 0.90 & 0.96 & 1.12 & 1.03 & 1 & 0.99 & 0.98 & 1.04 & 1.00 & 1.04 & 1.03 & 0.90 \\
2 & 0.97 & 0.91 & 1.10 & 0.96 & 0.99 & 0.99 & 0.99 & 2 & 1.02 & 0.99 & 1.10 & 0.90 & 1.02 & 0.99 & 1.05 \\
3 & 1.03 & 1.06 & 0.93 & 0.91 & 0.84 & 0.96 & 1.09 & 3 & 0.97 & 0.97 & 0.92 & 1.06 & 1.02 & 0.97 & 1.00 \\
4 & 1.04 & 0.94 & 1.04 & 0.92 & 0.99 & 1.07 & 0.93 & 4 & 1.01 & 1.01 & 1.02 & 1.04 & 0.93 & 1.01 & 1.03 
\enddata
\end{deluxetable*}

\section{MCMC fitting}\label{sec:mcmc}
To derive the total flux density ($S_{\rm tot}$), core size ($\theta_{\rm core}$), and relative position ($x, y$) of M60, we fitted the calibrated visibility data using the MCMC method, as recommended by \citet{Salafia2022}. The logarithmic likelihood equation used in the fitting is: 
\begin{eqnarray}
    \ln L(x) =-\frac{1}{2}\Sigma_{i=0}^{N} w_{i}[(V_{R,\,m}(u_{i},\,v_{i},\,\theta)-V_{R,\,i})^{2} +(V_{I,\,m}(u_{i},\,v_{i},\theta)-V_{I,i})^2]
\end{eqnarray}
where $V_{R,\,i}$ and $V_{I,\,i}$ represent the real and imaginary parts of the $i$'th visibility at the position $(u_{i},\,v_{i})$, and $w_{i}$ is data weight.  $V_{R,\,m}(u,\,v,\,\theta)$ and $V_{I,\,m}(u,\,v,\,\theta)$ are the real and imaginary parts of the model visibility, where $\theta=(S_{\rm tot},\,\theta_{\rm core},\,x,\,y)$ are parameters to be estimated. The model visibility function is defined as:
\begin{equation}
    V_{\rm m} = S_{\rm tot} \exp{\left[-2\pi^{2}\left(\frac{\theta_{\rm core}}{\sqrt{8\ln 2}}\right)^{2}(u^{2}+v^{2})-2\pi i (ux+vy)\right]}
\end{equation}
where $i=\sqrt{-1}$.

\begin{figure*}[ht!]
\includegraphics[width=0.49\textwidth]{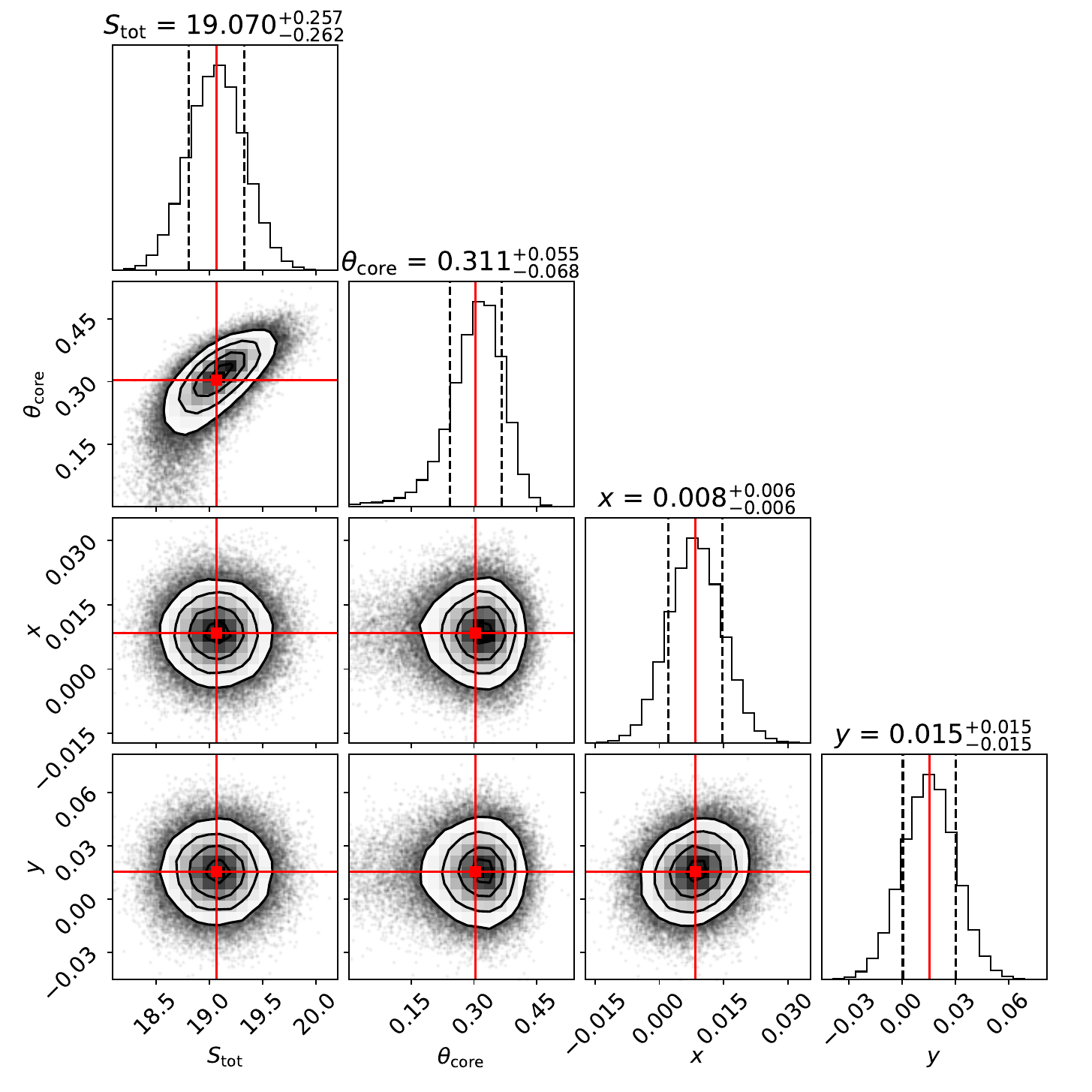}
\includegraphics[width=0.49\textwidth]{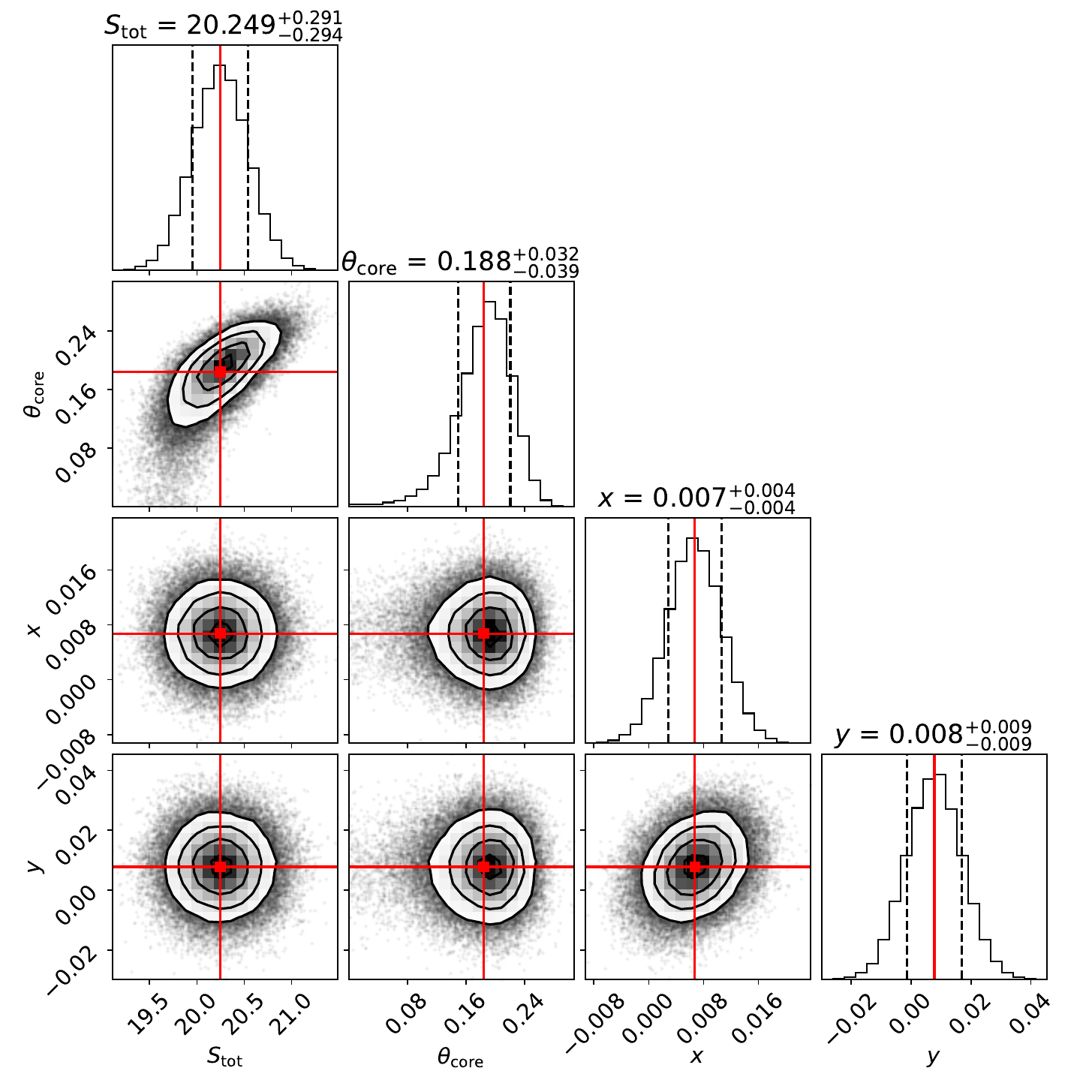}
\caption{Corner plots of the circular-Gaussian model fitting results for VLBA visibility data (Left for 4.4~GHz and Right for 7.6~GHz). There were 10, 000 MCMC simulations in the plot. The red lines represent the medians. The four parameters are total flux density in mJy,  core size in mas,  $x$ and $y$ coordinates in mas. \label{fig:corner}}
\end{figure*}

The MCMC fitting was executed using MCMCUVFit, a Python-based software specifically designed for fitting visibility data using MCMC method. The input model for MCMC fitting need four parameters: $S_{\rm tot}$, $\theta_{\rm core}$, $x$ and $y$. The search ranges for these four parameters to be estimated were set as follows: 10--25\,mJy for $S_{\rm tot}$, $10^{-6}$--2\, mas for $\theta_{\rm core}$, -1--1\, mas for both $x$ and $y$. The MCMC fitting process was conducted with 200 walkers and $10^{4}$ iterations. The initial half of the samples were discarded, and results are presented using corner plots in Figure \ref{fig:corner}. The parameters fitted through this process are listed in Table \ref{tab:model}.

\section{Lifetime of the dark cluster}\label{sec:lifetime}

Our approach to estimating the lifetime of the dark cluster present here follows \citet{Maoz1995, Maoz1998} and references therein. Since our objective is to exclude a dark cluster by illustrating that its lifetime is very short, we only consider the cluster that could survive for a long time. We presuppose that the structure of the cluster is described by Plummer model, with each constituent object assumed to have the same mass and zero temperature. 

The dark cluster could be composed of variety of objects, such as (1) black holes with $m_{*}\geq3~M_{\odot}$; (2) neutron stars with $1.4\leq m_{*} \leq 3~M_{\odot}$; (3) brown dwarfs and (4) white dwarfs with $10^{-3} \leq m_{*} \leq 1.4~M_{\odot}$. The lifespan of a cluster is determined by the mass and radius of the objects that constitute it. We use $R_{\rm S}$ to represent the black hole radius. The radius of a neutron star is given by the equation
\begin{equation}
    r_{*}(m_{*}) = \frac{(18\pi)^{2/3}}{10} \frac{\hbar^{2}}{Gm_{*}^{1/3}}\left(\frac{1}{m_{\rm H}}\right)^{8/3}
\end{equation}
where $\hbar$ is Plank constant and $m_{\rm H}$ is the mass of hydrogen atom. The radius of a white dwarf at zero temperature is expressed by the formula
\begin{equation}
    r_{*} (m_{*}) = \frac{1.57\times10^{9}}{\mu} \frac{[1-(m_{*}/M_{3})^{4/3}]^{1/2}}{(m_{*}/M_{3})^{1/2}} ~ {\rm cm}
\end{equation}
where $\mu$ is the mean molecular weight and $M_{3}=5.816\mu^{-2}~M_{\odot}$. Lastly, the radius of a brown dwarf is represented by
\begin{equation}
    r_{*}(m_{*}) = 2.2\times10^{9} \left(\frac{m_{*}}{M_{\odot}}\right)^{-1/3} \times \left[1+\left(\frac{m_{*}}{0.0032\,M_{\odot}}\right)^{-1/2}\right]^{-4/3}~{\rm cm}
\end{equation}

The lifetime of a dark cluster is constrained by two processes: evaporation and physical collisions. The maximum possible lifetime of the dark cluster depends on cluster's half-mass $M_{\rm h} \equiv M/2$ and its half-mass density $\rho_{\rm h}$, where $M$ is mass of the cluster. The evaporation process is a consequence of weak gravitational scattering. The evaporation timescale of a cluster can be described as follows:
\begin{equation}\label{eq:t_evap}
    t_{\rm evap} \simeq \frac{4.3\times10^{4}(M_{\rm h}/m_{*})}{\ln [0.8(M_{\rm h}/m_{*})]} \left(\frac{\rho_{\rm h}}{10^{8}M_{\odot}\,{\rm pc^{-3}}}\right)^{-1/2}~{\rm yr}
\end{equation}
The typical timescale for each object to physically collide with another is given by:
\begin{eqnarray}
\label{eq:t_coll}
    t_{\rm coll}(r_{*},r_{*}) =\nonumber \\ 
     \left[ 23.8 G^{1/2} M_{\rm h}^{1/3} \rho_{\rm h}^{7/6} \left(\frac{r_{*}^{2}}{m_{*}}\right)\left(1+\frac{m_{*}}{2^{1/2}\rho_{\rm h}^{1/3} M_{\rm h}^{2/3} r_{*}}\right) \right]^{-1} ~ {\rm s}
\end{eqnarray} 
For a given $M_{\rm h}$ and $\rho_{h}$, the lifetime of a cluster composed of objects with mass $m_{*}$ and radius $r_{*}$ is 
\begin{equation}
    \tau(r_{*},m_{*}) = \min\left[t_{\rm coll},t_{\rm evap}\right]
\end{equation}
The maximum lifetime of a dark cluster, $\tau_{\rm max}$, can be determined by examining all classes of dark objects mentioned above, across different mass ranges, i.e.,
\begin{equation}
    \tau_{\rm max}(M_{\rm h},\rho_{\rm h}) = \max\left[\tau(r_{*}, m_{*})\right]
\end{equation}
Figure \ref{fig:tau} show the $\tau_{\rm max}$ for each pair of $M_{\rm h}$ and $\rho_{\rm h}$.


\bibliography{M60ref}{}
\bibliographystyle{aasjournal}



\end{document}